# High throughput spatially sensitive single-shot quantitative phase microscopy

**Azeem Ahmad[1], Vishesh Dubey[1], Nikhil Jayakumar[1], Anowarul Habib[1], Ankit Butola[1], Mona Nystad[2], Ganesh Acharya[2,3], Purusotam Basnet[2], Dalip Singh Mehta[4], and Balpreet Singh Ahluwalia[1,3]**

[1]Department of Physics and Technology, UiT The Arctic University of Norway, Tromsø, Norway
[2]Department of Clinical Medicine, Women's Health and Perinatology Research Group, Faculty of Health Sciences, UiT - The Arctic University of Norway, Tromsø, Norway and Department of Obstetrics and Gynecology, University Hospital of North Norway, Tromsø, Norway
[3]Department of Clinical Science, Intervention and Technology Karolinska Institute, and Center for Fetal Medicine, Karolinska University Hospital, Stockholm, Sweden
[4]Applied Optics and Biophotonics Laboratory, Department of Physics, Indian Institute of Technology Delhi, Delhi, India

Email: ahmadazeem870@gmail.com and balpreet.singh.ahluwalia@uit.no

**Abstract**

High space-bandwidth product with high spatial phase sensitivity is indispensable for a single-shot quantitative phase microscopy (QPM) system. It opens avenue for widespread applications of QPM in the field of biomedical imaging. Temporally low coherence length light sources are generally implemented to achieve high spatial phase sensitivity in QPM at the cost of either reduced temporal resolution or smaller field of view (FOV). On the contrary, high temporal coherence light sources like lasers are capable of exploiting the full FOV of the QPM systems at the expense of less spatial phase sensitivity. In the present work, we employed pseudo-thermal light source (PTLS) in QPM which overcomes the limitations of conventional light sources. The capabilities of PTLS over conventional light sources are systematically studied and demonstrated on various test objects like USAF resolution chart and thin optical waveguide (height ~ 8 nm). The spatial phase sensitivity of QPM in case of PTLS is measured to be equivalent to that for white light source. The high-speed and large FOV capabilities of PTLS based QPM is demonstrated by high-speed imaging of live sperm cells that is limited by the camera speed and by imaging extra-ordinary large FOV phase imaging on histopathology placenta tissue samples.

## 1. Introduction

Quantitative phase microscopy (QPM) is a powerful technique which provides the measurement of different parameters of cells including their dynamics such as thickness and refractive index fluctuations. QPM is also used to determine cell's dry mass density, i.e., non-aqueous content [1]. High space-bandwidth product (SBP) with high spatial phase sensitivity is indispensable for QPM system. The SBP in QPM is defined as the product of the area of coherence limited field of view (FOV) and the area of the spatial frequency band of an interferometric image [2]. To achieve high SBP in single-shot QPM systems, a highly temporally and spatially coherent (i.e., laser) light source is generally utilized to obtain the interference pattern easily throughout the FOV of the camera [3, 4]. However, high temporal and spatial coherence (SC) properties of laser light source degrade the image quality due to coherent noise and parasitic fringe formation due to multiple reflections from different surfaces of the optical components [5]. As a consequence, it reduces the spatial phase sensitivity and height measurement accuracy of the system.

The spatial phase sensitivity of QPM can be improved by employing broadband light sources like white light (halogen lamp) and light emitting diodes (LEDs) [6-8]. These light sources provide high phase sensitivity due to their low temporal coherence (TC) length compared to narrow-band lasers. It is well known that the interference pattern occurs only when the optical path length difference between the object and the reference beam is within the coherence length of the light source [9, 10]. Therefore, it becomes difficult to obtain interference pattern quickly while employing low TC length light sources (coherence length $l_c \sim$ 2-6 μm). Moreover, high fringe density of the interference signal over the whole camera FOV cannot be obtained in case of low TC equipped off-axis QPM systems based on Linnik and Mach Zehnder interferometers. This overall reduces the SBP of QPM system due to coherence limited FOV. However, the SBP can be improved with on-axis digital holography which can utilize whole FOV of camera at the cost of low acquisition speed. On axis holography requires multiple interferograms for phase recovery of specimens [11-13]. This limits the ability to study the live cell dynamics of the biological cells like sperm and adds complexity to the system. In addition, the interferometric system becomes more sensitive to the external vibrations and introduce fringe like modulation error in the reconstructed phase maps [12]. Moreover, the use of spectrally broad band light sources in interferometry systems requires dispersion compensation and chromatic aberration corrected optical components [9, 10].

The high fringe density over the whole camera FOV can be achieved by coherence plane tilting even for broadband light sources [7, 14]. The optical configurations utilize diffraction grating into the beam path for coherence plane tilting [7, 14]. The white light diffraction phase microscopy (WL-DPM) is one of the techniques which works on the same principle to generate high fringe density with white light at the detector [14]. In WL-DPM, grating splits the input beam into several orders: zero, +1 and higher orders. One of the beams (say 0 order) is then spatially filtered to generate reference beam which interferes with the object beam (+1 order) to form modulated intensity patterns. To generate the reference beam, a pinhole of large diameter (~ 200 μm) is inserted into the beam path which does not completely block the object information and leads to the formation of halo in the recovered phase maps [14]. The imaging FOV in off-axis interference microscopy system has also been increased using τ interferometer in the past [15]. However, the recovered phase images suffered from the coherent noise generated due to the implementation of laser diode in the interferometer [3]. Thus, providing high sensitivity (<5 mrad) together with large FOV in a single-shot QPM is still unaccomplished.

Recently, a significant amount of work has been done to implement QPM with pseudo-thermal light source (PTLS) [16-19]. It is also called dynamic speckle pattern interferometry [20, 21]. Such light sources have been implemented in the quantitative phase imaging of optical waveguides and biological specimens previously [19-23]. The PTLS is generated by passing a highly temporal coherent laser beam through a rotating diffuser or a rough surface. The output of the diffuser acts as a highly temporally coherent (depends on the spectral bandwidth of the laser) and partially spatially coherent (depends on the source size) light source. PTLS thus carries the advantages of both broadband such as halogen lamp, LEDs in terms of high spatial phase sensitivity and narrowband such as laser sources in terms of the generation of high-density interference fringes over whole camera FOV and high photon degeneracy. In addition, PTLS does not suffer from chromatic aberration of the optical components and dispersion issues from the biological specimens as it has narrow temporal spectral bandwidth like laser. Here, we investigate the role of different light sources in QPM for enhancing the SBP. We demonstrate that PTLS can deliver single-shot phase imaging over large FOV without sacrificing the spatial phase sensitivity and the temporal resolution with imaging speed limited only by the camera acquisition speed.

In the present work, first, a comparison of different light sources white light (WL), filtered white light (FWL), laser and PTLS in terms of an optical image quality is systematically done. It is observed that PTLS generates speckle free images of USAF chart unlike in case of laser. The quality of the optical images is found to be comparable to the WL/FWL sources. Further, the spatial phase sensitivity of the QPM employed with FWL, laser, speckle fields generated by stationary diffuser and PTLS is compared. The phase sensitivity in case of PTLS is found to be comparable to the FWL source, which provides the maximum phase sensitivity in any QPM system [12, 13]. However, the short TC length of WL/FWL confines the high-density interference fringes (required for single shot QPM) in a small FOV of the detector. Contrary to this, PTLS can form high density interference fringe over the whole camera FOV with superior quality. The PTLS increases the SBP of the QPM system ~18 times compared to that for WL source. This makes the PTLS ideal for high speed, high-sensitivity quantitative phase imaging application. We have implemented PTLS-enabled QPM for single shot, high speed and highly spatial sensitive phase imaging of fast-moving sperm cells. In addition, large area FOV phase imaging is also demonstrated on placenta tissue sample.

## 2. Material and methods

### 2.1. Space bandwidth product

The SBP of an optical microscope is the product FOV and the area of spatial frequency information collected by the microscope and defined as follows [2]:

$$\text{SBP}_{Microscope} = \text{FOV} \times \pi \left(NA/\lambda\right)^2 \qquad (1)$$

In QPM, the FOV of the reconstructed phase maps is limited by the interference FOV. Therefore, the SBP of an interference microscope can be calculated by employing the following modified expression:

$$\text{SBP}_{QPM} = \text{Area of interference FOV} \times \pi \left(NA/\lambda\right)^2 \qquad (2)$$

Where, $\pi \left(NA/\lambda\right)^2$ is the area of the spatial frequency band, NA is the numerical aperture of the objective lens and $\lambda$ is the illumination wavelength.

### 2.2. Temporal coherence

In the theory of the optical fields, temporal coherence defines the correlation between the light vibrations at two different moments of time. According to Wiener- Khinchin theorem, temporal coherence function also called autocorrelation '$\Gamma(\Delta t) = \langle E(t)E^*(t - \Delta t)\rangle$' between the light fields is the Fourier transform of source power spectral density and can be expressed as follows:

$$\Gamma(\Delta t) = \int_{-\infty}^{\infty} S(v) \exp(i2\pi v \Delta t) \, dv \qquad (3)$$

Where, $S(v)$ is the source spectral distribution function, $\Gamma(\Delta t)$ is the temporal coherence function, and $\Delta t$ is the time delay between the optical fields $E(t)$ and $E^*(t - \Delta t)$.

The full width half maximum (FWHM) of temporal coherence function provides information about the temporal coherence length. Thus, larger spectral bandwidth light sources such as WL, FWL and LEDs etc. lead to smaller temporal coherence length or vice versa.

### 2.3. Spatial coherence

The spatial coherence governs the phase relationship between the light field vibrations reaching at two different points in space at the same instance of time. In transverse spatial coherence, the correlation function $\Gamma(\vec{r_1}, \vec{r_2}, \Delta t = 0) = \langle E(\vec{r_1}, t)E^*(\vec{r_2}, t)\rangle$ between the light fields originated from source $S(u, v)$ at two different spatial points $P_1(\vec{r_1}, z)$ and $P_2(\vec{r_2}, z)$ located in $(x, y)$ plane is considered. Analogous to Wiener- Khinchin theorem, the transverse spatial coherence function '$\Gamma(\vec{r_1}, \vec{r_2}, \Delta t = 0)$' and spatial angular frequency spectrum of the light source forms Fourier transform pairs. According to Van-Cittert-Zernike theorem, the correlation function can be defined as follows [24-26]:

$$\Gamma(\vec{r_1}, \vec{r_2}, \Delta t = 0) = \int_{-\infty}^{\infty} S(k_x, k_y) \exp[i(k_x \delta x + k_y \delta y)] \, dk_x dk_y \qquad (4)$$

Where, $S(k_x, k_y)$ is the source's angular spatial frequency spectrum, $\vec{r_1}$ and $\vec{r_2}$ are the position vectors of points $P_1$ and $P_2$ located in the plane of observation at points $(x, y)$ with respect to light source, $k_x$ and $k_y$ represents the spatial frequencies of the light field along $x$ and $y$ directions respectively.

For a plane wave propagating along the direction $\vec{N}(\cos \theta_x, \cos \theta_y)$, the spatial frequency components $k_x$ and $k_y$ can be given by the following expressions [24]:

$$k_x = \frac{2\pi}{\Lambda_x} = \frac{2\pi}{\lambda}\cos\theta_x = \frac{\omega}{c}\cos\theta_x \tag{5}$$

$$k_y = \frac{2\pi}{\Lambda_y} = \frac{2\pi}{\lambda}\cos\theta_y = \frac{\omega}{c}\cos\theta_y \tag{6}$$

where, $\Lambda_x$ and $\Lambda_y$ are the spatial periods of the light field along $x$ and $y$ directions. $\theta_x$ and $\theta_y$ are the angles of the propagating field direction from $x$ and $y$ axes, respectively, and $\lambda$ is the wavelength of light. The transverse spatial coherence lengths along $x$ and $y$ direction can be expressed as follows:

$$\rho_x = \frac{2\pi}{\Delta k_x}, \qquad \rho_y = \frac{2\pi}{\Delta k_y} \tag{7}$$

where, $\Delta k_x$ and $\Delta k_y$ are the transverse spatial frequency range along $x$ and $y$ axis.

The expressions for transverse coherence lengths of the optical field along $x$ and $y$ directions in terms of the angles can be defined as follows [24]:

$$\frac{1}{\rho_x} \approx \left(\frac{2}{\lambda} + \frac{1}{l_c}\right)\sin\theta_x, \qquad \frac{1}{\rho_y} \approx \left(\frac{2}{\lambda} + \frac{1}{l_c}\right)\sin\theta_y \tag{8}$$

where, $l_c$ is the temporal coherence length of the light source and given by $l_c \approx \lambda^2/\Delta\lambda$. $\Delta\lambda$ is the spectral bandwidth of the light source.

It can be seen from Eq. (8) that the transverse coherence of the source is governed by both the angular spatial frequency spectrum and temporal frequency spectrum of the optical field. The first and second terms in Eq. (8) represent the spatial and temporal coherence part of the light field, respectively. For WL and FWL, second term dominates over the first term and limits the confinement of interference fringes in a limited FOV due to low TC length. Whereas, PTLS has a narrow temporal spectral bandwidth depending on the laser's temporal spectral bandwidth and wide angular spatial frequency spectrum depending on the source size. In case of PTLS, confinement of the interference fringes depends on the spatial term rather than on the temporal term. The extremely wide angular spatial frequency spectrum (i.e., extended) light source leads to the generation of short spatial coherence length or vice versa and also limits the confinement of the fringes in a limited FOV. Thus, adequate source size of PTLS has to be kept for the formation of fringes over wider FOV.

### 2.4. Experimental setup

The experimental scheme of QPM system is illustrated in Fig. 1, which is based on the principle of non-common-path Linnik based interference microscopy. The light beam coming from a laser is splited into two beams using beam splitter $BS_1$. One of the beams directed towards microscope objective $MO_1$ which coupled the light beam into a single mode fiber (SMF) to generate spatially filtered temporally high and spatially high coherent light beam. The other one directed towards microscope objective $MO_2$ which illuminated the rotating diffuser (RD) with a diverging beam of spot size at diffuser plane ~ 1 – 2 mm. The RD generated temporally varying speckle field and leads to the reduction in speckle contrast significantly [27]. The scattered photons at the output of RD are coupled into a multi-mode fiber (MMF) with the help of beam splitter $BS_2$ and lens $L_1$. The core diameter of MMF is 1 mm. The RD followed by MMF generated uniform illumination, i.e., speckle free field, at the output port of MMF, which acts as an extended purely monochromatic light source namely PTLS. Thus, generated a temporally high and spatially low coherent light source corresponding to narrow-band laser. The MMF also carried the light coming from a halogen lamp as illustrated in Fig. 1. The output of MMF was both temporally and spatially incoherent for WL. To generate a filtered white light (FWL), a narrow bandpass filter of spectral bandwidth ~10 nm is inserted into the WL beam path after MMF output.

The output port of MMF is attached with the input port of interference microscopy system as shown in Fig. 1. The combination of lenses $L_2$ and $L_3$ relayed the source image (MMF output) at the back focal plane of the microscope objective MO$_3$ (10×/0.25NA) to achieve uniform illumination at the sample plane. The beam splitter BS$_3$ splitted the input beams into two; one is directed towards the sample and the other one towards reference mirror (M). The light beams reflected back from the sample and mirror recombined and formed interference pattern at the same beam splitter plane, which is projected at the camera plane with the help of L$_4$. The angle of reference mirror M controlled the angle between the object and the reference beam, i.e., the fringe width/density of the interferogram. The reference mirror M is kept at a particular angle for which high fringe density without aliasing effect is observed at Complementary Metal Oxide Semiconductor (CMOS) camera.

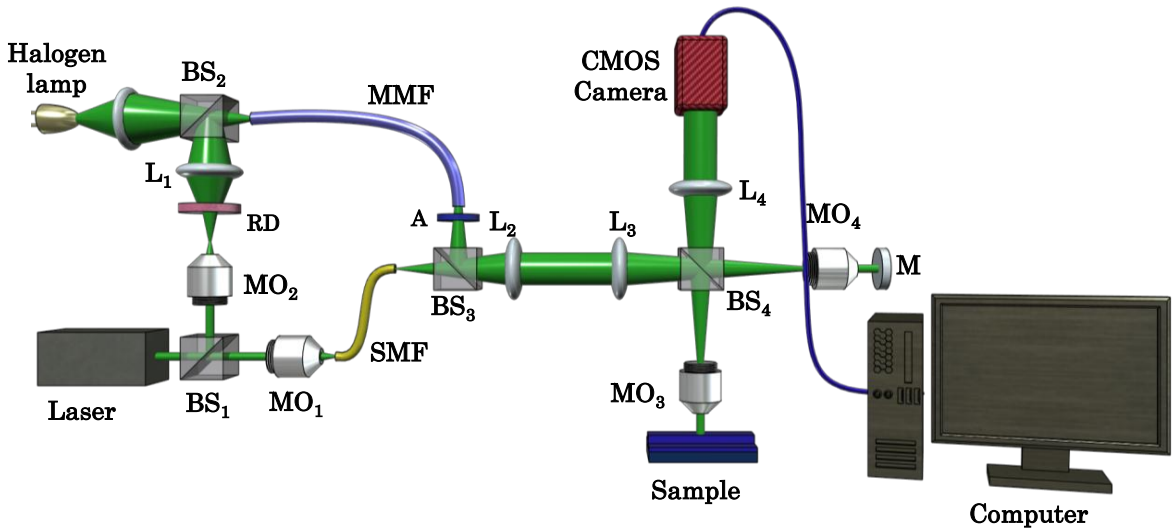

**Figure 1. Schematic diagram of the QPM system.** MO$_{1-4}$: Microscope objectives; BS$_{1-4}$: Beam splitters; L$_{1-4}$: Lenses; RD: Rotating diffuser; MMF: Multi-mode fiber; SMF: Single mode fiber; A: bandpass filter; M: Mirror and CMOS: Complementary metal oxide semiconductor camera.

## 2.5. Sample preparation

### 2.5.1. Sperm cells

The semen samples were collected according to the guidelines of WHO from men who attended the IVF clinic for the investigation/volunteers. The high motility sperm fraction were isolated using gradient centrifugation method. Complete details of the whole sample preparation can be found elsewhere [23]. Later, the sperm sample was diluted to a concentration of $5.0 \times 10^6$ cells/ml in culture medium. For QPM measurements the sperm cells were placed in a PDMS chamber on a reflective Si substrate and covered with standard #1.5 thickness coverslip. Informed consent was obtained from all participants and the ethical clearance was approved by The Regional Committee for Medical and Health Research Ethics of Norway (REK_nord).

### 2.5.2. Human placenta tissue sections

The placenta tissue sample was collected immediately after delivery, sectioned and rines in physiological salin to remove any contamination. Placental samples with size 1 mm$^3$ were immersed in 5 ml 8% formaldehyde in PHEM buffer and incubated at 4°C overnight. Tissue samples were immersed in 0.12% glycine at 37°C (1 h), infiltrated with 2.3 M sucrose at 4°C overnight to prevent freeze damage/structure deformation and transferred to liquid Nitrogen tank for storage. The samples were cut into 1 μm thick cryosections (EMUC6 ultramicrotome, Leica Microsystems, Vienna, Austria), and placed on Si substrate. Thereafter, samples were washed and mounted with PBS before imaging.

## 3. Results

### 3.1. Optical imaging of USAF chart

Most of the optical imaging systems frequently use different types of conventional light sources like halogen lamp, LEDs and lasers for the illumination of the sample. In bright field microscopy, halogen lamp or LED light sources being temporally low coherent are used to generate speckle free images of the specimens. The spatial coherence of such light sources is improved in the past by introducing a pinhole (50 – 100 μm) into the beam path at the expense of huge optical power loss from several W to mW. Thus, these light sources have very low photon degeneracy, i.e., average number of photons per unit coherence volume, compared to narrowband light sources like lasers [28]. High photon degeneracy of the lasers leads to formation of coherent and speckle noise in the images and significantly degrades their quality. Figures 2(a) and 2(b) illustrate the optical images of USAF resolution test target (Thorlabs: part #R3L3S1N) acquired using 10×/0.25NA objective lens corresponding to WL and FWL of 10 nm spectral bandwidth. A narrow band pass filter of peak wavelength 632.8 nm and spectral bandwidth of 10 nm (Thorlabs: part # FL632.8-10) is inserted into the WL beam path to generate FWL.

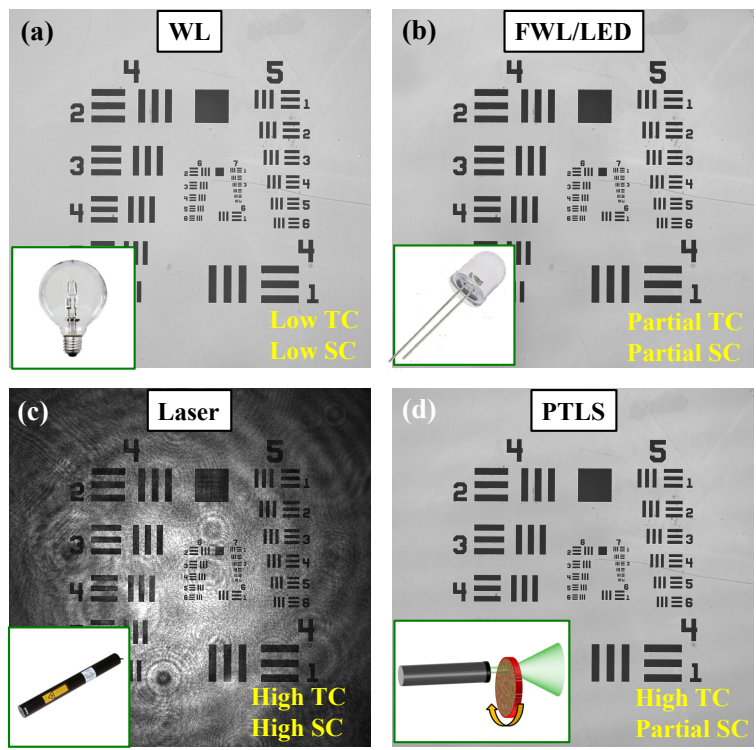

**Figure 2. Different light source for QPM.** Different light sources are compared in terms of the low coherent noise in non-interferometric images of USAF resolution chart. Low temporal coherent light sources like white light (WL) and filtered white light (FWL) produce noise free optical images, whereas direct laser generates coherent noise in the images. On the contrary, pseudo-thermal light source generates images comparable to the WL/FWL keeping the advantages of monochromatic light source.

It can be clearly visualized that the low TC light sources generate speckle free images of the specimens. Next, a narrowband laser Cobolt Flamenco™ (λ = 660 nm; Δλ = 0.001 pm) is used to illuminate the same sample. The optical image of the resolution chart for laser illumination is depicted in Fig. 2(c). The image clearly shows the unwanted parasitic fringe pattern noise which is not related to the test object. Thus, laser degrades the quality the images significantly. On the contrary, PTLS which has low SC and high TC properties, can improve the quality of the images comparable to WL or FWL light source. The microscopic image obtained by PTLS of the resolution chart is illustrated in Fig. 2(d). Thus, the implementation of PTLS in bright field optical microscopy outperforms the conventional light sources in terms of photon degeneracy (compared to WL and FWL) and coherent noise reduction (compared to lasers).

## 3.2. Spatial phase sensitivity comparison

Reduction of the coherent noise from the images by PTLS as compared to the direct laser is demonstrate above. In this section, the effect of the coherence properties of different light sources on the spatial phase sensitivity of QPM system is systematically studied. An optically flat silicon wafer is placed under the QPM system to record the interferometric images corresponding to FWL (bandwidth = 10 nm), direct laser, speckle field (stationary diffuser) and PTLS (rotating diffuser) as illustrated in Figs. 3(a), 3(d), 3(g) and 3(j), respectively. It can be clearly seen that FWL and PTLS generate coherent noise free interferograms. The high TC length of the laser forms unwanted interference fringes due to the superposition of multiple reflections coming from different surfaces of the optical components degrading the image quality. Conversely, PTLS equipped QPM generate coherent noise free interferograms like in case of FWL or LEDs. It can be visualized from Fig. 3(g), the interferogram is filled with the speckle noise when the diffuser is kept stationary. The size of each interferogram is 512 × 512 pixels.

The interferograms depicted in Figs. 3(a), 3(d), 3(g) and 3(j) are further post processed to recover corresponding phase maps using Fourier transform fringe analysis method [29]. The corresponding recovered phase images are illustrated in Figs. 3(b), 3(e), 3(h) and 3(k), respectively. Further, the effect of the coherent noise on the recovered phase images is calculated in terms of a quantitative matric called spatial phase sensitivity [12]. In order to measure the spatial phase sensitivity of the QPM system for FWL, laser, speckle field and PTLS, the standard deviation of the recovered phase maps is quantified. The standard deviation of the recovered phase maps is given in Table 1. The histogram plots corresponding to FWL, laser, speckle field and PTLS are presented in Figs. 3(c), 3(f), 3(i) and 3(l), respectively. It can be visualized that the spatial phase sensitivity of PTLS is comparable to the FWL/LEDs and much higher than that for laser. Thus, PTLS is capable to provide highly spatial phase sensitive images of the specimens while keeping a high TC length comparable to laser and subsequently enables the generation of high-density fringes over large FOV. PTLS provides ~17 times more spatial phase sensitivity compared to laser.

**Table 1:** Standard deviation calculated from the recovered phase maps corresponding to FWL, laser, speckle field and PTLS.

| S. No. | Light source | STD 'σ' (mrad) |
|---|---|---|
| 1. | FWL/LED | 2.6 |
| 2. | Laser | 59.4 |
| 3. | Speckle field | 66.5 |
| 4. | PTLS | 3.5 |

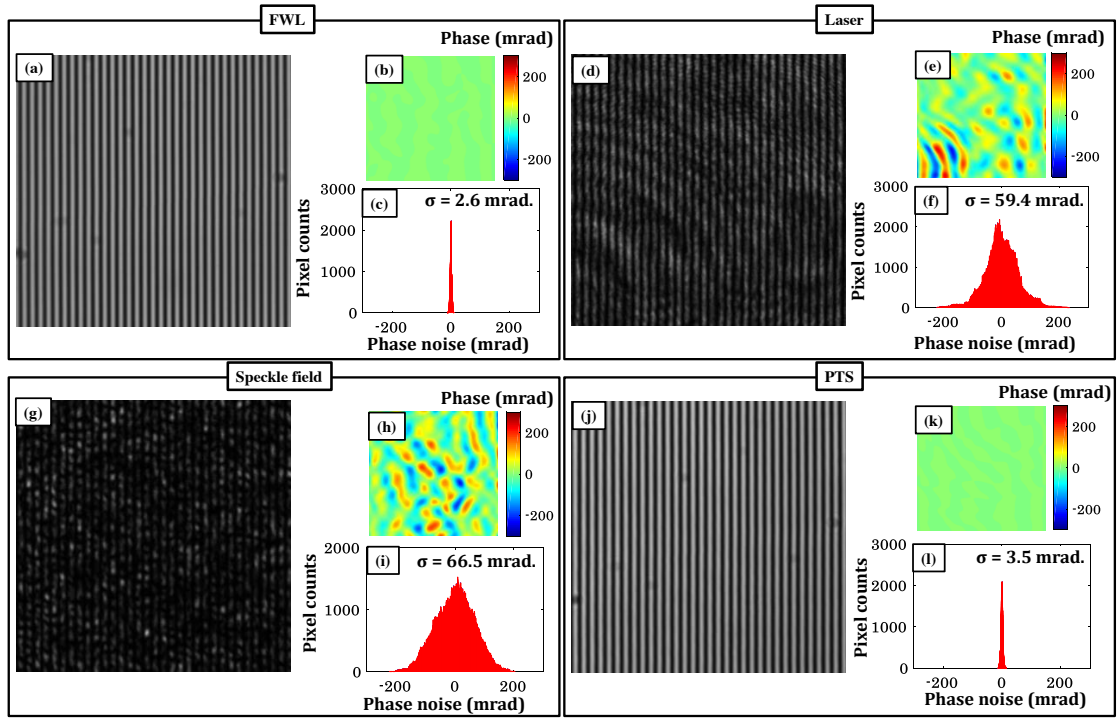

**Figure 3. PTLS prevents speckle noise formation.** Four differently generated light sources were used: filtered white light, narrow band laser, speckle field and pseudo-thermal light source. A bandpass filter of 10 nm bandwidth at central wavelength 632 nm is inserted into the white light beam path to generate FWL. Speckle field is generated by passing a narrow band laser beam through a stationary diffuser. PTLS is synthesized by passing a highly temporal coherent beam through a rotating diffuser followed by a multimode fiber bundle. (a, d, g, j) Interferometric images of a standard Si wafer. (b, e, h, k) Corresponding reconstructed phase images while employing FWL, laser, speckle field and PTLS, respectively. (c, f, I, l) Histograms of the recovered phase images of Si wafer to illustrate the spatial phase noise while using aforementioned light sources. The PTLS effectively suppresses spatial phase noise, behaving similarly to the FWL but very differently from conventional lasers. The color bars are in mrad.

### 3.3. Demonstration of high spatial phase sensitivity: Phase imaging of waveguide of 8 nm rib height

Next, the experiment is conducted on a rib optical waveguide having core material of silicon nitride ($Si_3N_4$), refractive index n ~ 2.04 and with a rib height of only ~ 8 nm. The fabrication procedure of the waveguide can be found elsewhere [30]. Here, a diode laser ($\lambda$ = 638 nm; $\Delta\lambda$ = 0.7 nm) is used instead of extremely high TC length Cobolt Flamenco™ laser. The diode laser is preferred over Cobalt laser as it generates less coherent noise due to short TC length. This makes a fair comparison between PTLS and diode laser in terms of the phase measurement sensitivity enhancement.

The rib waveguide is placed under the QPM (Fig. 1) to acquire high fringe density interferograms corresponding to direct laser and PTLS both as depicted in Figs. 4(a) and 4(e), respectively. It is depicted from Fig. 4(a), direct laser's interferogram suffers from the coherent noise generated due to high TC length of the laser. In contrast, PTLS produces coherent and speckle noise free interferogram as illustrated in Fig. 4(e). The insets of Figs. 4(a) and 4(e) clearly depict the fringe quality difference in the interferometric images generated from laser and PTLS. The recorded interferograms are further post-processed for the phase recovery of rib waveguide. Figures 4(b) and 4(f) illustrate the full FOV reconstructed phase maps of the rib waveguide corresponding to laser and PTLS, respectively. Figures 4(c) and 4(g) illustrate the zoomed views of the region marked with yellow dotted box in Figs. 4(b) and 4(f). Subsequently, the recovered phase maps can also be used to measure the height maps of rib waveguide as the refractive index of the material is known from ellipsometry experiments [31].

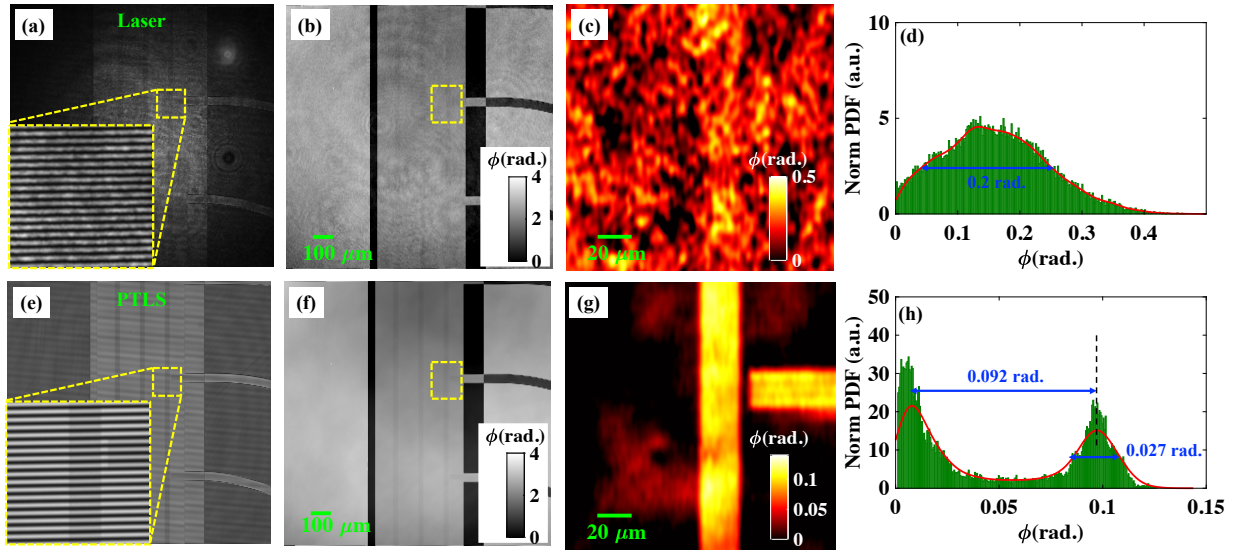

**Figure 4. PTLS produces high quality phase images of optical waveguide of 8 nm rib height with high resolution.** (a, e) Interferometric images of $Si_3N_4$ optical waveguide while using diode laser (at 638 nm) and PTLS, respectively. The insets depict the zoomed view of the region marked with yellow dotted box. (b, f) Reconstructed full FOV phase maps of an optical waveguide (H ~ 8 nm) corresponding to diode laser and PTLS, respectively. (c, g) Zoomed views of the regions marked with yellow dotted boxes in the reconstructed phase images depicted in Figs. 4(b) and 4(f), respectively. (d, h) The histogram plots of the phase images shown in Figs. 4(c) and 4(g), respectively. It can be visualized that the PTLS produces superior quality phase images of the waveguide compared to diode laser. The thin rib height of ~ 8 nm is not visible and embedded in the coherent noise of the QPM system in case of diode laser. The color bars are in rad.

The histogram plots of the recovered phase maps are depicted in Figs. 4(d) and 4(h), respectively. It can be observed from the histogram plots that laser generates large coherent noise in the recovered phase maps which is significantly very less in case of PTLS. In PTLS histogram plot, the peaks are well resolved, whereas the peaks are not distinct in case of laser. The phase information of rib waveguide is completely embedded in the phase noise for laser. The difference between the peaks shown in Fig. 4(h) is equal to the difference between the waveguide's foreground and background phase values. The bandwidth of the peaks, i.e., standard deviation exhibits the phase variation over the waveguide surface. The FWHM of the peak is measured to be equal to 0.027 rad. The little large value of standard deviation could be due to following possible reasons: optical thickness variation of the waveguide, phase noise and optical aberration present in the QPM system. It is quite evident from Figs. 4(d) and 4(h) that the coherence property of the light source plays an important role in the phase measurement sensitivity of the QPM system. Thus, PTLS provides approximately 10-fold increase in the spatial phase sensitivity compared to diode laser.

### 3.4. Demonstration of high space bandwidth product: Interferometric imaging of USAF chart

Here, we demonstrate the advantages of PTLS over the WL/FWL source in terms of enhancing the SBP. For this, experiments are conducted on USAF resolution test target by sequentially employing WL, FWL, laser and PTLS in QPM (Fig. 1). The interferograms of USAF chart corresponding to WL, FWL, laser and PTLS are sequentially recorded using developed QPM system as depicted in Figs. 5(a), 5(d), 5(g) and 5(j), respectively. Figures 5(b), 5(e), 5(h) and 5(k) represents the zoomed views of the regions marked with green color boxes depicted in Figs. 5(a), 5(d), 5(g) and 5(j), respectively. The insets of Figs. 5(b), 5(e), 5(h) and 5(k) represent the line profiles of the interferograms along red dotted lines corresponding to WL and PTLS. The line profiles of the whole camera FOV corresponding to WL, FWL, laser and PTLS along blue dotted lines are illustrated in Figs. 5(c), 5(f), 5(i) and 5(l), respectively.

It is evident from Fig. 5(c) that the interference fringes are confined only in a limited region of the camera FOV for WL source. This is due to the low TC length of WL source. The area of the interference FOV can be increased by inserting a narrow spectral band-pass filter (Δλ = 10 nm) into the beam path at the expense of large intensity loss. It can be visualized that narrow band-pass filter of 10 nm spectral bandwidth is not sufficient to produce interference fringe over whole camera sensor. On the contrary, the interference fringes of the same quality as that for WL/FWL are observed over the whole camera FOV while using PTLS as illustrated in Fig. 5(k). The area of the interference FOV for WL is found to be 18 times less than that for PTLS. The normalized interference fringe contrast value below 0.1 is used to select the boundary of the interference FOV. Thus, the implementation of the PTLS increases the SBP of QPM system by 18-fold compared to WL, without sacrificing the temporal resolution. It can be seen that direct laser can obviously forms the high-density fringes over whole camera FOV at the cost of large coherent noise (Fig. 5i). In addition, the fringe visibility is not constant over whole camera FOV. This could be due to the gaussian intensity profile of the light beam coming out from a single mode fiber. On the contrary, the line profile exhibited in Fig. 5(l) clearly demonstrates that the fringe visibility does not vary over whole camera FOV in case of PTLS.

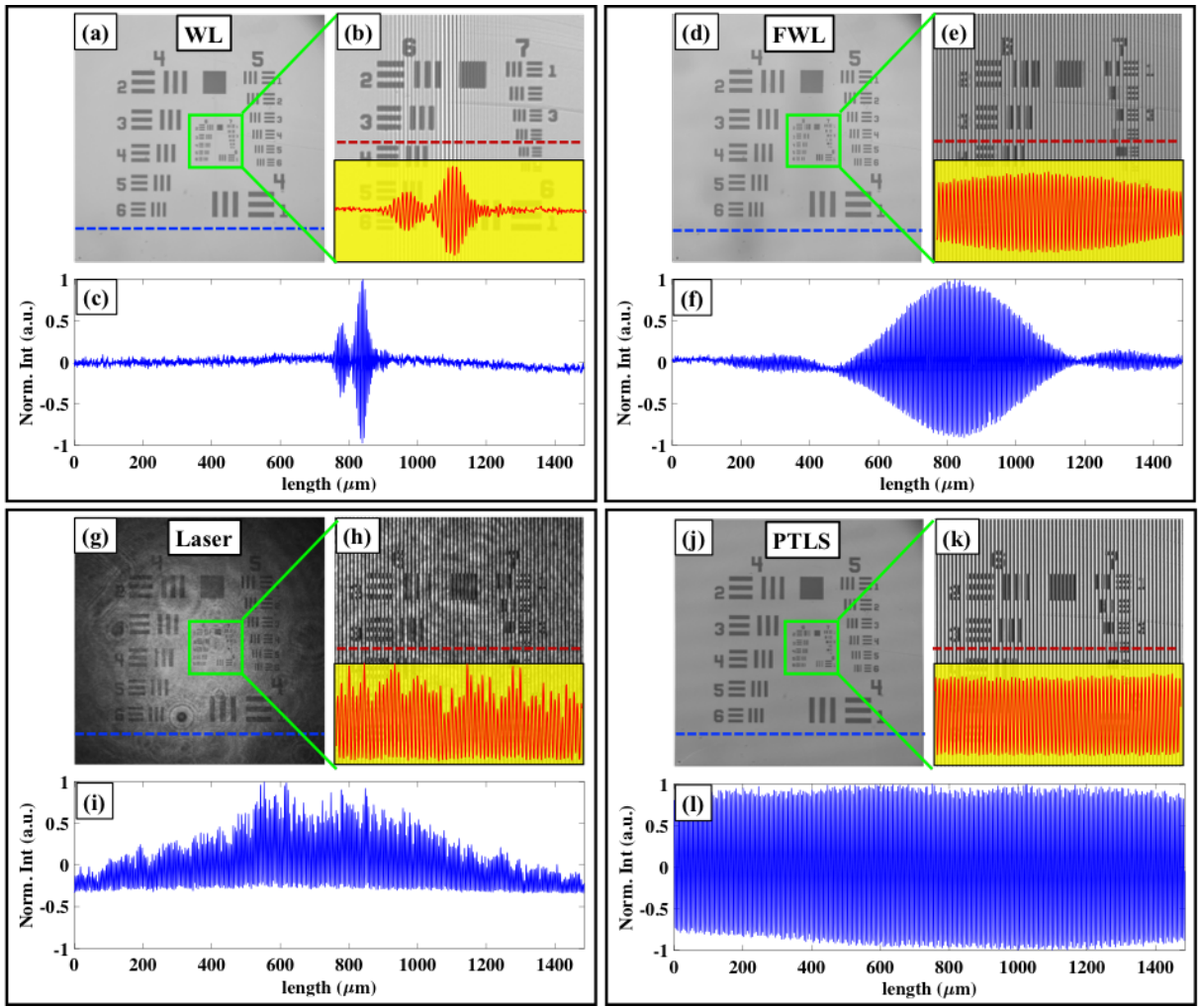

**Figure 5. PTLS generates high space bandwidth product interferometric images without speckle noise.** Four light sources with different degrees of temporal and spatial coherence were used: a white light (WL), filtered white light (FWL), narrow band laser and pseudo-thermal light source (PLTS). (a, d, g, j) Interferometric images of USAF resolution chart while using WL, FWL, Laser and PTLS, respectively. (b, e, h, k) Zoomed view of the region marked with green color box in Figs. 5a, 5d, 5g and 5j, respectively. The insets depict the line profiles along red dotted horizontal lines. (c, f, i, l) The line profiles over the whole camera FOV along blue dotted lines shown in Figs. 5a, 5d, 5g and 5j, respectively. PTLS produces coherent noise free interference fringes over the whole camera FOV unlike WL/FWL and narrowband laser.

## 3.5. Demonstration of high-speed QPM with high spatial phase sensitive: QPM of fast-moving living human sperm cells

Next, the experiment is conducted on the live human sperm cells using water immersion (60×/1.2NA) objective lens. The reference objective is not changed and kept 10×/0.25NA. Live human sperm cells are always challenging to image in their natural environment because of the following reasons: the mean swimming speed of sperm cells is 20-160 µm/s [32-34] and the tail of the sperm is very thin and generates minute optical thickness [35]. To image healthy sperm cells, the QPM system should be capable to acquire highly spatial sensitive interferometric/phase images at very high acquisition speed more than 120 fps. As mentioned in the previous sections, the implementation of PTLS in QPM system is capable to provide highly spatial sensitive phase images at very high speed.

To prepare the sperm sample for phase imaging, first, a 170 µm thick polydimethylsiloxane (PDMS) chamber is placed on top a reflecting silicon wafer. The diluted sperm sample's volume of 10 – 15 µl is placed in a ~ 4 mm × 4 mm opening area of the PDMS. The sample is then covered from the top using a cover glass (thickness: 170 µm) to avoid any air current in the sample. The sample is placed under the QPM system to perform high speed and high-resolution phase imaging of live human sperm cells. The interferometric images of the sperm cells are acquired at 30 fps, which is the maximum speed of our camera at full resolution. This can be obviously improved by using a high speed camera such as 200 – 1000 fps, supported by camera pco.dimax HS4, Phantom T1340 [36].

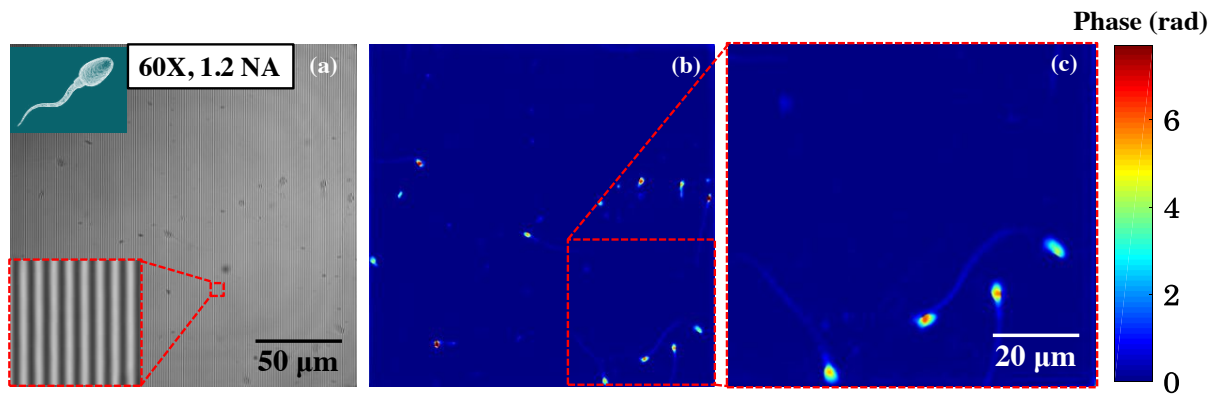

**Figure 6. PTLS provides high resolution video rate phase images of fast moving transparent biological cells.** A water immersion objective lens with 60×/1.2 NA is used to image fast moving human sperm cells. (a) Interferometric image of live human sperm cells (see Visualization 1). The inset depicts the zoomed view of the region marked with red dotted box. (b) Reconstructed high-resolution phase image of the sperm cells (Visualization 2). (c) The zoomed view of the phase image of sperm cells enclosed by red dotted box shown in Fig. 6b. The scale bars are in micron. The color bar is in rad.

The interferometric image of the live sperm is depicted in Fig. 6(a). The inset represents the zoomed view of the region marked with red dotted box. The live interferometric movie of the live sperm cells can be seen through Visualization 1. The reconstructed phase map of one interferometric frame of the movie is illustrated in Fig. 6(b). The phase movie of the live sperm cells can be found in Visualization 2. The zoomed view of the region marked with red dotted box in Fig. 6(b) is shown in Fig 6(c). Due to high sensitivity, we can also visualize the phase map of thin tail region of the sperm cells which is around 100 nm [35]. Thus, PTLS enables both high-speed imaging with high spatial sensitive quantitative phase imaging. Thanks to the high-density fringes formed over large FOV, single-shot PTLS based QPM systems can be deliver quantitative phase maps of thin biological samples moving at fast speeds.

## 3.6. Demonstration of large FOV together with high spatial phase sensitive: QPM of placenta tissue sample

To demonstrate the large FOV phase imaging capability of PTLS equipped QPM system, the experiments are conducted on human placenta tissue samples of 1 µm thickness. The images are acquired with 20×/0.45 NA objective lens which provides FOV approximately equal to 660 × 660 µm$^2$.

The sequentially acquired interferometric images of the whole tissue sample are illustrated in Fig. 7(a) and 7(b). The sample stage has motorized translation along X and Y axis to acquire the interferometric images of side-by-side FOV. The overlap between the images is kept approximately equal to 20%. It is demonstrated that with just two interferometric images acquired in 500 msecs are sufficient to cover the whole tissue samples as shown in Fig. 7. The interferograms are further utilized to recover the phase maps of the tissue sample. Figure 7(c) exhibits the stitched phase image of the whole tissue sample. The total FOV of the stitched image is 660 × 1200 µm$^2$. The FOV is just limited by the camera chip size and the magnification of the objective lens employed, while the temporal resolution (imaging speed) is limited by the camera sensor speed.

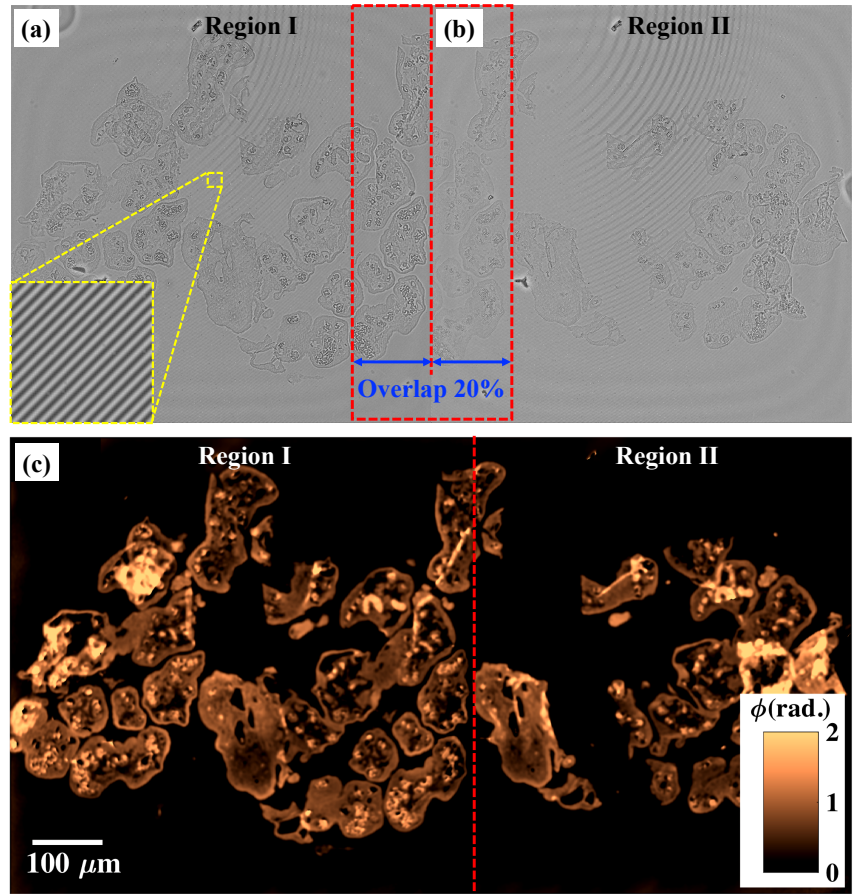

**Figure 7. Large FOV phase imaging.** (a, b) Interferometric images of the placenta tissue sample. The inset depicts the zoomed view of a small region of the interferogram marked with yellow dotted box. (c) Stitched phase image of the whole tissue sample of 1 µm thickness. The color bar is in rad.

## 4. Discussion and conclusion

PTLS offers many folds increase in SBP compared to WL source without sacrificing the spatial phase sensitivity and temporal resolution. In addition, PTLS enables the use of single-shot phase recovery algorithms without introducing coherent noise in the recovered phase images which is otherwise a common problem while working with lasers. The comparison between different light sources, in terms of coherent noise, spatial phase sensitivity and SBP, is demonstrated on USAF chart and on an optical waveguide with rib height of 8 nm. It is observed that PTLS outperforms the use of conventional light sources like WL, FWL, LEDs and Lasers in QPM system. Being, a single shot technique video rate quantitative phase imaging is successfully demonstrated on live human sperm cells, with speed only limited to the camera speed. This became possible due to the presence of high-density fringes over the whole camera FOV, which enabled single shot phase recovery of the fast-moving sperm cells. Thus, the FOV obtained by PTLS based QPM is just limited by the camera chip size and the magnification of the objective lens employed, while the temporal resolution (imaging speed) is limited by the camera sensor speed.

Moreover, QPM equipped with laser light sources on evaluating of cells for clinical application are not considered as appropriate, particularly in IVF clinic for the evaluation of sperm cells and embryo quality. Therefore, QPM data for evaluating sperm cell quality with PTLS has a potential to be applied in IVF clinic successfully. The concept of large FOV phase imaging capability of PTLS based QPM system is also demonstrated on placenta tissue samples. In pathology, generally a large area of sample sections has to be scanned to provide a holistic understanding about diseases and their progression. We believe that the capabilities of PTLS paves the way to achieve its wider penetration in the field of quantitative phase imaging in life sciences application and would possibly also find applications in metrology. The present technique may find applications in the high speed QPM and can be integrated with other functionalities like optical waveguide trapping [3, 37], microfluidics [38], optical tweezers [39] and multi-modal optical imaging techniques [40].

**Funding**

BSA acknowledges the funding from the INPART project, Research Council of Norway, (project # NANO 2021–288565), (project # BIOTEK 2021–285571).

**Acknowledgements**

The authors would like to acknowledge Randi Olsen for cryosectioning of placenta tissue sections.

**Author Contribution**

A.A. has conceptualized the idea, designed and performed most of the experiments, and analyzed the data. A.A. and B.S.A. mainly wrote the manuscript. V. D. prepared the biological samples, acquired the data, and performed data processing. N.J. A.H., A.B. assisted during all the experiments. M.N., G.A., and P.B. provided the biological specimens. B.S.A., D.S.M., P.B. and G.A. conceived the project and supervised this work. All authors reviewed and edited the manuscript.

**Disclosures**

The authors declare no conflicts of interest.